\documentclass[twocolumn,pre,aps,nofootinbib]{revtex4}

\usepackage{graphicx}
\graphicspath{{./fig/}}

\usepackage{bm}

\begin{document}


%

\newcommand{\EQ}{\begin{equation}}
\newcommand{\EN}{\end{equation}}
\newcommand{\EQA}{\begin{eqnarray}}
\newcommand{\ENA}{\end{eqnarray}}
\newcommand{\eq}[1]{(\ref{#1})}
\newcommand{\EEq}[1]{Equation~(\ref{#1})}
\newcommand{\Eq}[1]{Eq.~(\ref{#1})}
\newcommand{\Eqs}[2]{Eqs.~(\ref{#1}) and~(\ref{#2})}
\newcommand{\eqs}[2]{(\ref{#1}) and~(\ref{#2})}
\newcommand{\Eqss}[2]{Eqs~(\ref{#1})--(\ref{#2})}
\newcommand{\Sec}[1]{Sec.~\ref{#1}}
\newcommand{\Secs}[2]{Secs.~\ref{#1} and~\ref{#2}}
\newcommand{\App}[1]{Appendix~\ref{#1}}
\newcommand{\Fig}[1]{Fig.~\ref{#1}}
\newcommand{\FFig}[1]{Figure~\ref{#1}}
\newcommand{\Tab}[1]{Table~\ref{#1}}
\newcommand{\Figs}[2]{Figures~\ref{#1} and \ref{#2}}
\newcommand{\Tabs}[2]{Tables~\ref{#1} and \ref{#2}}
\newcommand{\bra}[1]{\langle #1\rangle}
\newcommand{\bbra}[1]{\left\langle #1\right\rangle}
\newcommand{\mean}[1]{\overline #1}
\newcommand{\mod}[1]{\mid\!\!#1\!\!\mid}
\newcommand{\chk}[1]{[{\em check: #1}]}

\newcommand{\Reynolds}{\mathrm{Re}}
\newcommand{\St}{\mathrm{St}}
\newcommand{\Da}{\mathrm{Da}}
\newcommand{\Sc}{\mathrm{Sc}}
\newcommand{\Pe}{\mathrm{Pe}}
\newcommand{\Ka}{\mathrm{Ka}}
\newcommand{\Rm}{\mathrm{Re}_{\rm M}}
\newcommand{\Pm}{\mathrm{Pr}_{\rm M}}
\newcommand{\ii}{\mathrm{i}}
\newcommand{\D}{D}

%
%
\newcommand{\gggg}{\bm{g}}
\newcommand{\ddd}{\bm{d}}
\newcommand{\rrr}{\bm{r}}
\newcommand{\xx}{\bm{x}}
\newcommand{\yy}{\bm{y}}
\newcommand{\zzz}{\bm{z}}
\newcommand{\uu}{\bm{u}}
\newcommand{\vv}{\bm{v}}
\newcommand{\ww}{\bm{w}}
\newcommand{\mm}{\bm{m}}
\newcommand{\PP}{\bm{P}}
\newcommand{\QQ}{\bm{Q}}
\newcommand{\UU}{\bm{U}}
\newcommand{\UUc}{\bm{U}_{\rm c}}
\newcommand{\bb}{\bm{b}}
\newcommand{\qq}{\bm{q}}
\newcommand{\BB}{\bm{B}}
\newcommand{\HH}{\bm{H}}
\newcommand{\II}{\bm{I}}
\newcommand{\AAA}{\bm{A}}
\newcommand{\aaa}{\bm{a}}
\newcommand{\aaaa}{\bm{a}} 
\newcommand{\eee}{\bm{e}}
\newcommand{\jj}{\bm{j}}
\newcommand{\JJ}{\bm{J}}
\newcommand{\nn}{\bm{n}}
\newcommand{\ee}{\bm{e}}
\newcommand{\ff}{\bm{f}}
\newcommand{\EE}{\bm{E}}
\newcommand{\FF}{\bm{F}}
\newcommand{\TT}{\bm{T}}
\newcommand{\CC}{\bm{C}}
\newcommand{\KK}{\bm{K}}
\newcommand{\MM}{\bm{M}}
\newcommand{\GG}{\bm{G}}
\newcommand{\kk}{\bm{k}}
\newcommand{\SSS}{\bm{S}}
\newcommand{\grav}{\bm{g}}
\newcommand{\nab}{\bm{\nabla}}
\newcommand{\OO}{\bm{\Omega}}
\newcommand{\oo}{\bm{\omega}}
\newcommand{\LL}{\bm{\Lambda}}
\newcommand{\llambda}{\bm{\lambda}}
\newcommand{\pomega}{\bm{\varpi}}
\newcommand{\nnn}{\hat{\mbox{\boldmath $n$}} {}}
%
%
\newcommand{\meanB}{\overline{B}}
\newcommand{\meanC}{\overline{C}}
\newcommand{\meanUU}{\overline{\bm{U}}}
\newcommand{\meanHHH}{\overline{\cal H}}
\newcommand{\meanFFFF}{\overline{\mbox{\boldmath ${\cal F}$}}{}}{}
%
%
\def\const{\rm const}
\def\Dc{D_{\rm c}}
\def\Dt{D_{\rm t}}
\def\DT{D_{\rm T}}
\def\tauc{\tau_{\rm c}}
\def\taueff{\tau_{\rm eff}}
\def\uoneD{{v'}}
\def\urms{u_{\rm rms}}
\def\cs{c_{\rm s}}
\def\kf{k_{\rm t}}
\def\sT{s_{\rm T}}
\def\sL{s_{\rm L}}
\def\vF{s_{\rm T}}
%
%
\newcommand{\RRRR}{\bm{\mathsf{R}}}
\newcommand{\SSSS}{\bm{\mathsf{S}}}
\newcommand{\IIII}{\bm{\mathsf{I}}}

\newcommand{\DD}{\mathrm{D}}
\newcommand{\dd}{\mathrm{d}}

\newcommand{\half}{{\textstyle{\frac{1}{2}}}}
\newcommand{\onethird}{{\textstyle{\frac{1}{3}}}}
\newcommand{\W}{\,{\rm W}}
\newcommand{\V}{\,{\rm V}}
\newcommand{\kV}{\,{\rm kV}}
\newcommand{\T}{\,{\rm T}}
\newcommand{\G}{\,{\rm G}}
\newcommand{\Hz}{\,{\rm Hz}}
\newcommand{\kHz}{\,{\rm kHz}}
\newcommand{\kG}{\,{\rm kG}}
\newcommand{\K}{\,{\rm K}}
\newcommand{\g}{\,{\rm g}}
\newcommand{\s}{\,{\rm s}}
\newcommand{\ms}{\,{\rm ms}}
\newcommand{\cm}{\,{\rm cm}}
\newcommand{\m}{\,{\rm m}}
\newcommand{\km}{\,{\rm km}}
\newcommand{\kms}{\,{\rm km/s}}
\newcommand{\kg}{\,{\rm kg}}
\newcommand{\Mm}{\,{\rm Mm}}
\newcommand{\pc}{\,{\rm pc}}
\newcommand{\kpc}{\,{\rm kpc}}
\newcommand{\yr}{\,{\rm yr}}
\newcommand{\Myr}{\,{\rm Myr}}
\newcommand{\Gyr}{\,{\rm Gyr}}
\newcommand{\erg}{\,{\rm erg}}
\newcommand{\mol}{\,{\rm mol}}
\newcommand{\dyn}{\,{\rm dyn}}
\newcommand{\J}{\,{\rm J}}
\newcommand{\RM}{\,{\rm RM}}
\newcommand{\EM}{\,{\rm EM}}
\newcommand{\AU}{\,{\rm AU}}
\newcommand{\A}{\,{\rm A}}
%
%
\newcommand{\yan}[3]{, Astron. Nachr. {\bf #2}, #3 (#1).}
\newcommand{\yact}[3]{, Acta Astron. {\bf #2}, #3 (#1).}
\newcommand{\yana}[3]{, Astron. Astrophys. {\bf #2}, #3 (#1).}
\newcommand{\yanas}[3]{, Astron. Astrophys. Suppl. {\bf #2}, #3 (#1).}
\newcommand{\yanal}[3]{, Astron. Astrophys. Lett. {\bf #2}, #3 (#1).}
\newcommand{\yass}[3]{, Astrophys. Spa. Sci. {\bf #2}, #3 (#1).}
\newcommand{\ysci}[3]{, Science {\bf #2}, #3 (#1).}
\newcommand{\ysph}[3]{, Solar Phys. {\bf #2}, #3 (#1).}
\newcommand{\yjetp}[3]{, Sov. Phys. JETP {\bf #2}, #3 (#1).}
\newcommand{\yspd}[3]{, Sov. Phys. Dokl. {\bf #2}, #3 (#1).}
\newcommand{\ysov}[3]{, Sov. Astron. {\bf #2}, #3 (#1).}
\newcommand{\ysovl}[3]{, Sov. Astron. Letters {\bf #2}, #3 (#1).}
\newcommand{\ymn}[3]{, Monthly Notices Roy. Astron. Soc. {\bf #2}, #3 (#1).}
\newcommand{\yqjras}[3]{, Quart. J. Roy. Astron. Soc. {\bf #2}, #3 (#1).}
\newcommand{\ynat}[3]{, Nature {\bf #2}, #3 (#1).}
\newcommand{\sjfm}[2]{, J. Fluid Mech., submitted (#1).}
\newcommand{\pjfm}[2]{, J. Fluid Mech., in press (#1).}
\newcommand{\yjfm}[3]{, J. Fluid Mech. {\bf #2}, #3 (#1).}
\newcommand{\ypepi}[3]{, Phys. Earth Planet. Int. {\bf #2}, #3 (#1).}
\newcommand{\ypr}[3]{, Phys.\ Rev.\ {\bf #2}, #3 (#1).}
\newcommand{\yprl}[3]{, Phys.\ Rev.\ Lett.\ {\bf #2}, #3 (#1).}
\newcommand{\yepl}[3]{, Europhys. Lett. {\bf #2}, #3 (#1).}
\newcommand{\pcsf}[2]{, Chaos, Solitons \& Fractals, in press (#1).}
\newcommand{\ycsf}[3]{, Chaos, Solitons \& Fractals{\bf #2}, #3 (#1).}
\newcommand{\yprs}[3]{, Proc. Roy. Soc. Lond. {\bf #2}, #3 (#1).}
\newcommand{\yptrs}[3]{, Phil. Trans. Roy. Soc. {\bf #2}, #3 (#1).}
\newcommand{\yjcp}[3]{, J. Comp. Phys. {\bf #2}, #3 (#1).}
\newcommand{\yjgr}[3]{, J. Geophys. Res. {\bf #2}, #3 (#1).}
\newcommand{\ygrl}[3]{, Geophys. Res. Lett. {\bf #2}, #3 (#1).}
\newcommand{\yobs}[3]{, Observatory {\bf #2}, #3 (#1).}
\newcommand{\yaj}[3]{, Astronom. J. {\bf #2}, #3 (#1).}
\newcommand{\yapj}[3]{, Astrophys. J. {\bf #2}, #3 (#1).}
\newcommand{\yapjs}[3]{, Astrophys. J. Suppl. {\bf #2}, #3 (#1).}
\newcommand{\yapjl}[3]{, Astrophys. J. {\bf #2}, #3 (#1).}
\newcommand{\ypp}[3]{, Phys. Plasmas {\bf #2}, #3 (#1).}
\newcommand{\ypasj}[3]{, Publ. Astron. Soc. Japan {\bf #2}, #3 (#1).}
\newcommand{\ypac}[3]{, Publ. Astron. Soc. Pacific {\bf #2}, #3 (#1).}
\newcommand{\yannr}[3]{, Ann. Rev. Astron. Astrophys. {\bf #2}, #3 (#1).}
\newcommand{\yanar}[3]{, Astron. Astrophys. Rev. {\bf #2}, #3 (#1).}
\newcommand{\yanf}[3]{, Ann. Rev. Fluid Dyn. {\bf #2}, #3 (#1).}
\newcommand{\ypf}[3]{, Phys. Fluids {\bf #2}, #3 (#1).}
\newcommand{\yphy}[3]{, Physica {\bf #2}, #3 (#1).}
\newcommand{\ygafd}[3]{, Geophys. Astrophys. Fluid Dynam. {\bf #2}, #3 (#1).}
\newcommand{\yzfa}[3]{, Zeitschr. f. Astrophys. {\bf #2}, #3 (#1).}
\newcommand{\yptp}[3]{, Progr. Theor. Phys. {\bf #2}, #3 (#1).}
\newcommand{\yjour}[4]{, #2 {\bf #3}, #4 (#1).}
\newcommand{\pjour}[3]{, #2, in press (#1).}
\newcommand{\sjour}[3]{, #2, submitted (#1).}
\newcommand{\yprep}[2]{, #2, preprint (#1).}
\newcommand{\pproc}[3]{, (ed. #2), #3 (#1) (to appear).}
\newcommand{\yproc}[4]{, (ed. #3), pp. #2. #4 (#1).}
\newcommand{\ybook}[3]{, {\em #2}. #3 (#1).}

\preprint{NORDITA-2010-42}

\title{Turbulent front speed in the Fisher equation: dependence on Damk\"ohler number}

\author{Axel Brandenburg}
   \affiliation{NORDITA, AlbaNova University Center, Roslagstullsbacken 23,
   SE-10691 Stockholm, Sweden}
   \affiliation{Department of Astronomy, Stockholm University,
   SE 10691 Stockholm, Sweden}
   \email{brandenb@nordita.org}
\author{Nils Erland L.\ Haugen}
  \affiliation{Sintef Energy Research, N-7034 Trondheim, Norway}
  \email{nils.haugen@phys.ntnu.no}
\author{Natalia Babkovskaia}
\email{NBabkovskaia@gmail.com}
  \affiliation{Division of Geophysics and Astronomy (P.O. Box 64),
   FI-00014 University of Helsinki, Finland}

\date{\today,~ $ $Revision: 1.75 $ $}

\begin{abstract}
Direct numerical simulations and mean-field theory are used to model
reactive front propagation in a turbulent medium.
In the mean-field approach, memory effects of turbulent diffusion are
taken into account to estimate the front speed in cases when the
Damk\"ohler number is large.
This effect is found to saturate the front speed to values comparable
with the speed of the turbulent motions.
By comparing with direct numerical simulations, it is found that the
effective correlation time is much shorter than for non-reacting flows.
The nonlinearity of the reaction term is found to make the front
speed slightly faster.
\end{abstract}
\maketitle

\section{Introduction}

It is well known that the propagation speed of a flame front is greatly
enhanced if a mixture of fuel and oxygen is in a turbulent state.
This topic of turbulent premixed combustion was pioneered by
Damk\"{o}hler \cite{D40} some 70 years ago and is reviewed extensively
in recent literature \cite{Pet99,BPBD05,Dri08}.
In spite of its importance, the question of burning velocities in a
turbulent medium continues to be of major importance even today
\cite{Kido02,VL06,Kitagawa08}.

Much of the current work is based on the original Damk\"{o}hler paradigm for
premixed combustion.
He distinguishes two regimes, namely those of large-scale and small-scale
turbulence.
For the purpose of the present paper it is useful to base this
distinction on a comparison of the mean turbulent flame width
with the scale of the energy-carrying eddies \cite{Pet99}.
In the small-scale turbulence regime, also referred to as the distributed 
reaction zone regime, the turbulent flame speed is computed
using a formula where the microscopic diffusivity is replaced by the sum of
microscopic and turbulent diffusivities.
This is possible because there is good scale separation.
This implies that the turbulent front thickness (i.e.\ the thickness of
the flame brush) is much broader than the scale of the turbulent eddies.
This regime is characterized by small Damk\"{o}hler numbers.
In the opposite case of large Damk\"{o}hler numbers, the turbulent
front thickness is smaller than the scale of the turbulent eddies and
can therefore no longer be described by turbulent diffusion.
This regime is characterized as that of large-scale turbulence.
In this case the turbulent front speed reaches its maximal value that
is given by the rms velocity of the turbulence in the direction of
front propagation.

The regime of large-scale turbulence is subdivided further into regimes
of corrugated and wrinkled flamelets, depending essentially on the ratio
of Reynolds number to Damk\"{o}hler number, which is also related to the
Karlovitz number.
When the Reynolds number is small compared with the Damk\"{o}hler number
(small Karlovitz number), the flame front is merely wrinkled, but for
large Reynolds numbers (large Karlovitz number) it becomes corrugated
and can consist of isolated flamelets detached from other parts of the front.
In the present paper we will mainly be concerned with the flame speed rather
than the question of whether the flame front is wrinkled or corrugated.

\begin{figure}[t!]\begin{center}\includegraphics[width=\columnwidth]{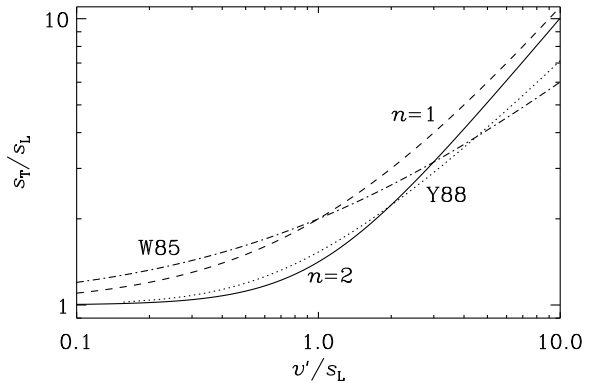}
\end{center}\caption[]{
Comparison of different expressions for the normalized front speed, $\sT/\sL$,
as a function of the turbulent velocity, $\uoneD/\sL$.
The labels $n=1$ and $n=2$ refer to \Eqs{n=1}{n=n}, while
Y88 and W85 refer to \Eqs{Y88}{W85}.
}\label{pyakhot}\end{figure}

In turbulent combustion, the averaged flame speed, $\sT$, is usually
normalized by the corresponding laminar flame speed, $\sL$, and one
is interested in the dependence on the normalized turbulent velocity,
$\uoneD$.
For the regime of large-scale turbulence, the speed-up ratio of turbulent to
laminar flame speed is given by the geometric increase of the wrinkled
surface area of the flame front.
Damk\"{o}hler assumed that the increase in surface area is proportional
to the ratio of the turbulent velocity of the eddies to the laminar
flame speed.
This leads to the expectation that the dependence of $\sT$ on
$\uoneD$ is given by \cite{Pet99}
\EQ
\sT/\sL=1+\uoneD/\sL.
\label{n=1}
\EN
This equations captures the expected limiting cases
that $\sT$ should not become larger than $\uoneD$ and that $\sT=\sL$
in the absence of turbulence, i.e., for $\uoneD=0$.
However, unsatisfactory agreement with measurements motivated the search
for other dependencies.
For example, Pocheau \cite{Poc92} derives the more general formula
\EQ
\sT/\sL=\left[1+\left(\uoneD/\sL\right)^n\right]^{1/n},
\label{n=n}
\EN
where $n$ is a parameter.
This formula obeys the aforementioned limiting
case for any value of $n$.
Pocheau \cite{Poc92} contrasts the formula with another one
proposed by Yakhot \cite{Yak88},
\EQ
\sT/\sL=\exp\left[\left(\uoneD/\sT\right)^2\right],
\label{Y88}
\EN
where $\sT<\uoneD$ for $\uoneD\to\infty$.
Yet another fit formula is given by
\EQ
\sT/\sL=1+C_{\rm W}\left(\uoneD/\sL\right)^m
\label{W85}
\EN
with fit parameters $C_{\rm W}$ and $m=0.7$ \cite{Wil85}.
Both \eq{Y88} and \eq{W85} have a front speed less than $v'$
for $\uoneD\to\infty$, provided $m<1$ in \Eq{W85}.
As can be seen from \Fig{pyakhot}, the different proposals for
the front speed are quite similar, making it difficult to use measurements
to distinguish between them.
Furthermore, realistic descriptions of flame properties are hampered by
the fact that feedback on the flow by the actual combustion process
depends to the specific case and is not easy to model.
The feedback on the flow is therefore usually ignored.
It might therefore be useful to return to a simple model of front
propagation that can be treated in more detail and to address the
unsettled question regarding the different proposals in
\Eqss{n=1}{W85} for the dependence of $\sT$ on $\uoneD$.
Following Kerstein \cite{Ker02}, we consider here the Fisher equation,
which is also known as the Kolmogorov--Petrovskii--Piskunov (KPP)
equation \cite{KPP37}.
An important difference to earlier work is the fact that we solve this
equation in the three-dimensional case in the presence of a turbulent
velocity field.

The Fisher or KPP equation is a simple scalar equation
that possesses propagating front solutions.
This equation is familiar in biomathematics \cite{Mur93} as a simple model
for the spreading of diseases.
It has also been amended by an advection term to describe the
interaction with a turbulent velocity field in one \cite{BN09}
and multiple \cite{PBNT10} dimensions,
the effects of cellular flows \cite{CTVV03}, and the scaling of the front
thickness \cite{BVV08}.
Furthermore the equation has also been modified to account for 
different interacting species,
that can be used to model the spreading of auto-catalytically polymerizing
left and right handed nucleotides \cite{BM04}.
Given that $C$ is a passive (albeit reacting) scalar, the Fisher equation
does ignore any feedback on the flow and is therefore well suited to help
clarifying questions regarding the relation between $\sT$ and $\uoneD$.

In the present paper we consider both direct numerical simulations (DNS)
of this equation in three dimensions as well as its averaged form
where the effects of turbulence are being parameterized by a non-Fickian
diffusion equation.
Such an equation allows for the ballistic spreading of a passive scalar
concentration on short time scales, which is expected to be important
when the front propagates at a speed comparable to that of the turbulence
itself.

\section{The Fisher equation}

A simple model of front propagation is the Fisher equation which,
in the simplest case, is a one-dimensional partial differential equation
\cite{Mur93,KPP37,Fis37,Kal84},
\EQ
\frac{\partial C}{\partial t}
={C\over\tauc}\left(1-{C\over C_0}\right)
+D\frac{\partial^2 C}{\partial x^2},
\label{dCdt}
\EN
for the concentration $C$.
Here, $\tauc$ is the chemical reaction time, $D$ is the diffusivity,
and $C_0$ is some saturation value above which further growth is quenched.
\EEq{dCdt} corresponds to an autocatalytic reaction were a reactant ${\cal R}$
yields a product ${\cal P}$ at a rate $k$ that is itself proportional to the
concentration of the products, $[{\cal P}]$, i.e.\
\EQ
{\cal R}\stackrel{k~}{\longrightarrow}{\cal P}\quad\mbox{with}\quad
k=[{\cal P}]/\tauc C_0.
\EN
This can then also be written as ${\cal P}+{\cal R}\to2{\cal P}$.
Saturation of the product concentration, $C=[{\cal P}]$,
results from the fact that the total mass is conserved,
i.e.\ $[{\cal R}]+[{\cal P}]=C_0=\const$.
The evolution equation for the concentration $C=[{\cal P}]$
is then given by \Eq{dCdt}.

This equation has two solutions, an unstable solution, $C=0$, and a stable
one, $C=C_0$.
The diffusion term seeds the neighboring regions that are in an unstable
state, which promotes the rapid transition from $C=0$ to $C=C_0$.
This leads to the propagation of the transition front in the direction
down the gradient of $C$ with a front speed \cite{Mur93}
\EQ
\sL=2\sqrt{D/\tauc},
\label{vF}
\EN
where the subscript L refers to the {\it laminar} front speed.

In many cases of practical interest the diffusion coefficient $D$
is rather small and is hardly relevant when there is rapid advection through
fluid motions.
In that case the governing equations become advection--reaction--diffusion
equations.
This can be written as
\EQ
\frac{\partial C}{\partial t}+\nab\cdot(\UU C)
={C\over\tauc}\left(1-{C\over C_0}\right)
+D\nabla^2C,
\label{dCdt_3D}
\EN
where $\UU$ is the flow speed.
If the flow is turbulent and has zero mean, there can be circumstances
where the average concentration $\meanC$ can be described by an
equation similar to \Eq{dCdt}, but with $C$ being replaced by the
mean value $\meanC$,
and $D$ being replaced by some turbulent diffusivity $\Dt$, i.e.\
\EQ
\frac{\partial\meanC}{\partial t}
={\meanC\over\tauc}\left(1-{\meanC\over C_0}\right)
+\DT\frac{\partial^2\meanC}{\partial x^2},
\label{dmeanCdt}
\EN
where $\DT=D+\Dt$ is the total (i.e.\ the sum of microscopic and
turbulent) diffusivity.
We have here assumed that the mean concentration shows a systematic
variation in the $x$ direction and have thus assumed averaging over
the $y$ and $z$ directions, so $\meanC(x,t)$ can be described
by a one-dimensional evolution equation.

Given the similarity between \Eqs{dCdt}{dmeanCdt},
one would expect that in the turbulent case with appropriate
initial conditions the effective turbulent propagation speed $\sT$
of the front can still be described by an expression similar to \Eq{vF},
but with $D$ being replaced by $\DT$, i.e.\
$\vF=2\sqrt{\DT/\tauc}$.
A useful estimate for the turbulent diffusivity is $\Dt=\urms/3\kf$,
where $\kf$ is the wavenumber of the energy-carrying eddies
and $\urms$ is the rms velocity of the turbulence \cite{BSV09}.
Thus, for $\Dt\gg D$, the effective value of $\vF$ is expected
to be $2(\urms/3\tauc\kf)^{1/2}$.
On the other hand, one cannot expect the front speed to increase
inde\-finitely with decreasing $\tauc$.
Indeed, one would not expect $\vF$ to exceed the rms velocity of the
turbulence in the direction of front propagation.
Following common practice, we denote it by $\uoneD$.
Under the assumption of isotropy, $\uoneD$ is related to the
three-dimensional rms velocity by $\uoneD=\urms/\sqrt{3}$.

An important nondimensional measure of $\tauc$ is the
Damk\"ohler number, which is the ratio of the turnover time,
$(\urms\kf)^{-1}$, to $\tauc$.
This number is here defined as
\EQ
\Da=(\tauc\urms\kf)^{-1}.
\label{Da}
\EN
Note that our definition of Da is based on the {\it wavenumber} $\kf$
rather than the {\it scale} $2\pi/\kf$, which would have reduced
the numerical value of Da by a factor of $2\pi$.
For small values of $\Da$ we expect $\vF\approx2\uoneD\,\Da^{1/2}$,
while for large values one expects $\vF\approx\uoneD$ \cite{Poc92}.
Thus, a more general formula is expected to be
\EQ
\vF^2=\uoneD^2f(\Da),
\EN
where $f(\Da)$ increases linearly with $\Da$ for $\Da\ll1$ and
$f(\Da)\approx1$ for $\Da\gg1$.
This saturation behavior can also be interpreted as a reduction of
the {\it effective} value of $\tauc$ \cite{EKR98}.
An important goal of this paper is to determine the form of the function
$f(\Da)$.

\section{Non-Fickian diffusion}

The Fickian diffusion approximation made in \Eq{dmeanCdt} for the mean
concentration $\meanC$ becomes invalid if $\meanC$ varies rapidly in time,
and in principle also in space.
This is indeed expected to be the case when $\Da\gg1$.
For rapid time variations,
\Eq{dmeanCdt} attains then an extra time derivative and takes the form
\cite{BKM04}
\EQ
\tau\frac{\partial^2\meanC}{\partial t^2}
+\frac{\partial\meanC}{\partial t}
={\meanC\over\tauc}\left(1-{\meanC\over C_0}\right)
+\DT\frac{\partial^2\meanC}{\partial x^2},
\label{dmeanC2dt2}
\EN
which is a damped wave equation with relaxation time $\tau$
and an additional reaction term.
The presence of the nonlinearity in the reaction term leads to an additional
contribution in the $\meanC$ equation which has here been ignored (see
\App{MeanField} for a more consistent treatment).

Without the reaction term,
\Eq{dmeanC2dt2} is also known as the telegraph equation.
This equation emerges naturally when computing turbulent transport
coefficients using the $\tau$ approximation \cite{BF03}.
Evidence for the existence of the wave term has been found from
isotropic forced turbulence simulations \cite{BKM04}.
A non-dimensional measure of $\tau$ is given by the Strouhal number,
\EQ
\St=\tau\urms\kf=\tau\urms^2/3\Dt,
\EN
where the first equality is useful for turbulence simulations
where $\tau\urms\kf$ is readily evaluated,
while the second equality is useful for the mean-field model, where $\kf$
does not appear explicitly and $\Dt$ and $\urms$ are given.

Using DNS of forced turbulence with a passive scalar, the value
of St has been determined to be around 3 by relating triple corrections
to quadratic ones \cite{BKM04}.
Although we consider the value of St as being fairly well constrained,
we do consider below a range of different values.

The purpose of this section is to study solutions of \Eq{dmeanC2dt2}
that can then be compared with DNS of the Fisher equation
coupled with the Navier-Stokes
equations for obtaining a turbulent velocity that enters \Eq{dCdt_3D}.
We consider first the case where $D$ is negligible
and solve \Eq{dmeanC2dt2} for different values of Da in a one-dimensional
domain that was chosen long enough so that the front speed can be
determined accurately enough.
We use a numerical scheme that is second order in space and third order
in time \cite{BD02}.
In some cases a resolution of $2^{15}\approx3\times10^4$ mesh points was
necessary.

\begin{figure}[t!]\begin{center}\includegraphics[width=\columnwidth]{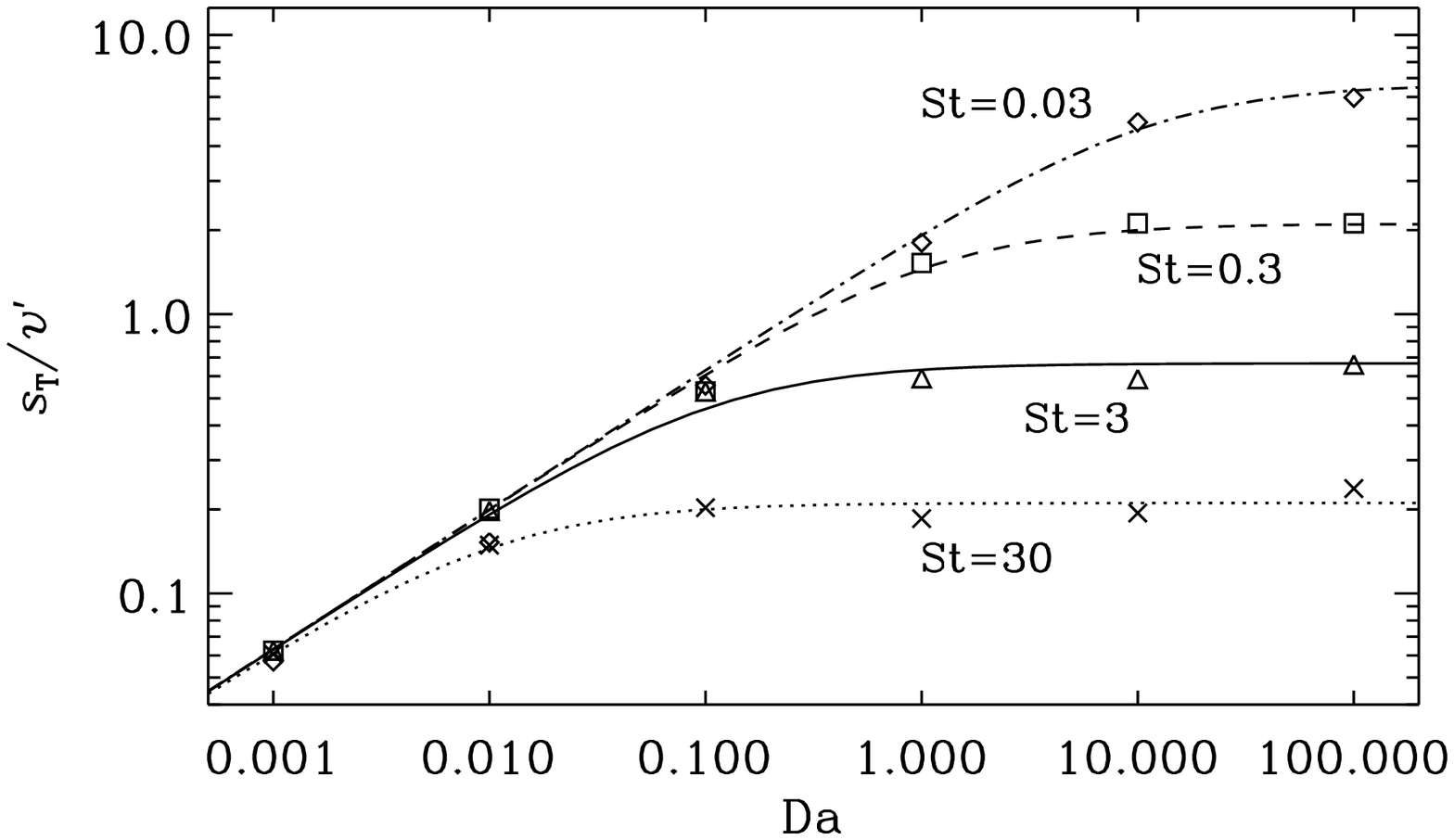}
\end{center}\caption[]{
Dependence of the front speed of solutions of \Eq{dmeanC2dt2}
on Da for different values of St and $\Pe=\infty$.
The lines represent fits given by \Eq{fit}.
}\label{pDa}\end{figure}

We study first the dependence of the front speed on Da for a range of
different values of St and $D\ll\Dt$.
Here, $\vF$ is determined by differentiating the concentration integrated
over the whole domain,
\EQ
\vF(t)={\dd\over\dd t}\int {\meanC\over C_0}\,\dd z,
\label{vFmeasure}
\EN
and approximating the asymptotic front speed with the value at the time
when the front has reached the other end of the domain.
This quantity is also known as the reaction speed.
This is indicated by $\meanC$  reaching a small fraction (e.g.\ $10^{-6}$)
of $C_0$.
The result is shown in \Fig{pDa}.
For small values of Da, the front speed is independent of the value of St
and we reproduce the anticipated result, i.e.\ $f(\Da)=4\,\Da$.
For larger values of Da the front speed reaches eventually a constant
value.
However, the limiting value depends on St.
Our results are well reproduced by the fit formula
\EQ
{\vF^2\over\uoneD^2}
\equiv f(\Da,\,\St)\approx{4\,\Da\over1+3\,\St\,\Da}.
\label{fit}
\EN
This formula obeys the anticipated limiting behaviors for small and large
values of Da, provided $\St\approx4/3$.
The numerically determined data agree quite well with \Eq{fit}.
However, in some cases the numerical data are somewhat uncertain and depend
also slightly on resolution and domain size.

\begin{figure}[t!]\begin{center}\includegraphics[width=\columnwidth]{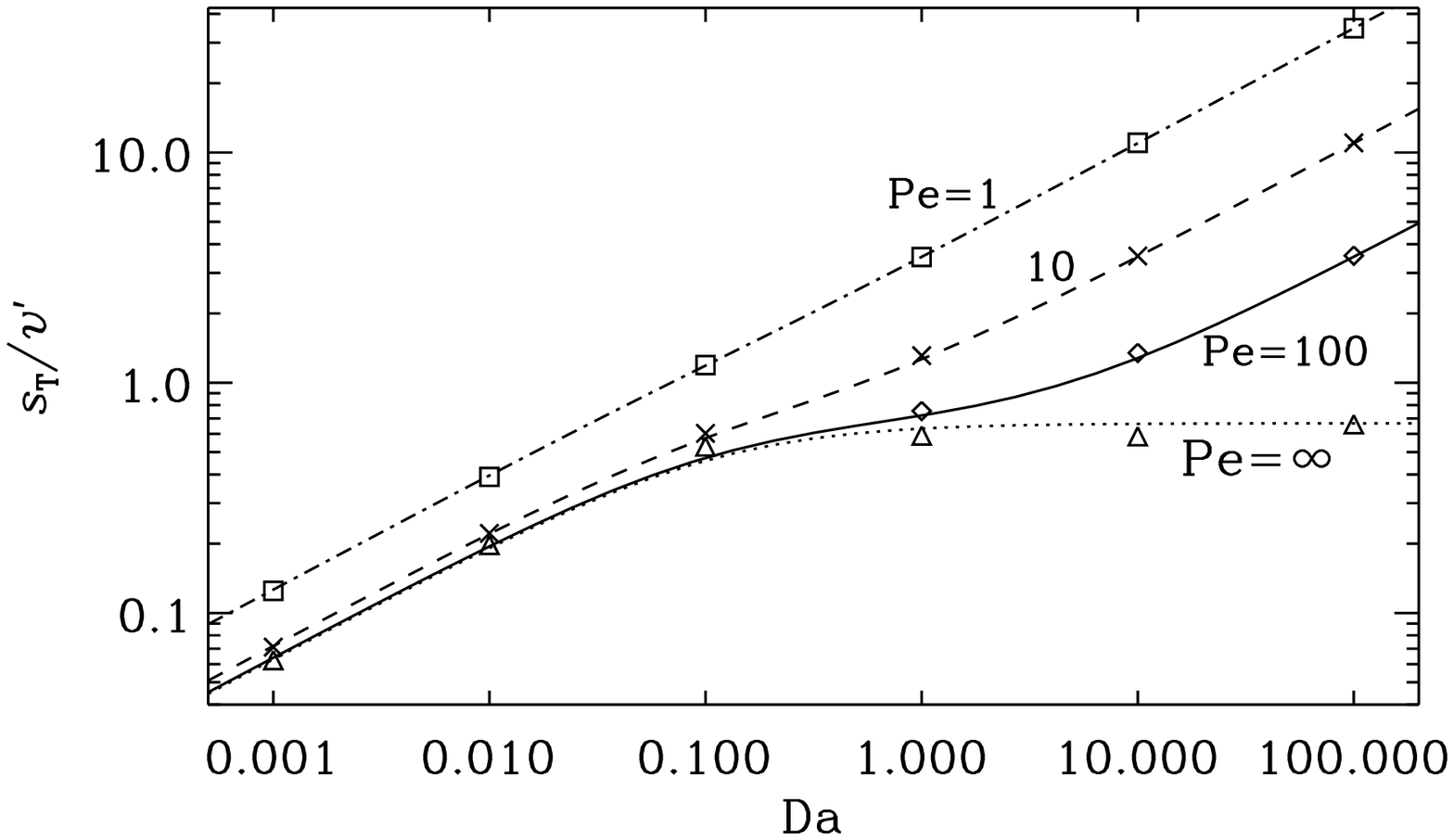}
\end{center}\caption[]{
Dependence of the front speed of solutions of \Eq{dmeanC2dt2}
on Da for different values of Pe and $\St=3$.
The lines represent fits given by \Eq{fitPe}.
}\label{pDaPe}\end{figure}

In the DNS presented below, where the numerical resolution is still limited,
the value of $D$ is often not negligible.
Its value is characterized by the Peclet number,
\EQ
\Pe=\urms/D\kf\approx3\Dt/D.
\label{Pe}
\EN
For small values of Da, the expression for the front speed
should be $\vF=2\sqrt{(D_{\rm t}+D)/\tauc}$,
which can then be written as
\EQ
\vF/\uoneD=2\sqrt{\left(1+3\Pe^{-1}\right)\Da}
\quad\mbox{(for $\Da\ll1$)}.
\EN
For larger values of $\Da$ we solve \Eq{dmeanC2dt2} numerically.
A good fit formula for $f$ is given by
\EQ
f(\Da,\,\St,\,\Pe)\approx4\Da
\left({3\over\Pe}+{1\over1+3\St\,\Da}\right).
\label{fitPe}
\EN
In \Fig{pDaPe} we compare the fit formula with the numerically obtained
front speeds for different values of Pe, keeping $\St=3$.
The agreement is again quite good.

\begin{figure}[t!]\begin{center}\includegraphics[width=\columnwidth]{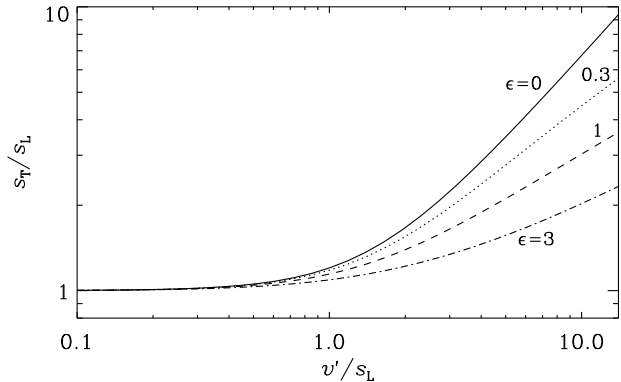}
\end{center}\caption[]{
Dependence of $\sT/\sL$ and $\uoneD/\sL$
for $\epsilon=0$ (solid line), 0.3 (dotted), 1 (dashed), and 3 (dash-dotted).
Note that there is no unique relationship between $\sT/\sL$ and $\uoneD/\sL$.
}\label{pst_urms}\end{figure}

In turbulent combustion it is customary to plot the normalized front speed,
$\sT/\sL$, as a function of the normalized turbulent velocity, $\uoneD/\sL$.
Using our definitions of $\Da$ and $\Pe$ in \Eqs{Da}{Pe}, respectively,
we have $\uoneD/\sL=({\Pe/12\,\Da})^{1/2}$ and find
\EQ
{\sT\over\sL}=\left[1+{1\over3\St/4+\epsilon\uoneD/\sL}
\left({\uoneD\over\sL}\right)^2\right]^{1/2},
\label{sT_Fisher}
\EN
where we have defined $\epsilon=\kf\ell_{\rm F}$ with
$\ell_{\rm F}=(\tauc D/12)^{1/2}$ being a measure for the
laminar flame thickness.
Note that $\epsilon$ can also be expressed in terms of
Da and Pe via
\EQ
\epsilon=(12\Da\,\Pe)^{-1/2}.
\EN
A more familiar quantity is the ratio $\ell/\ell_{\rm F}=\epsilon^{-1}$,
where $\ell=\kf^{-1}$ is the typical eddy scale.
Even if we can assume the value of St to be given,
$\epsilon$ is not a fixed quantity.
It is therefore clear that there cannot be a unique relationship
between $\sT/\sL$ and $\uoneD/\sL$.
Instead, there must be a family of solutions depending on
the value of $\epsilon$; see \Fig{pst_urms}.

\begin{figure*}[t!]\begin{center}\includegraphics[width=\textwidth]{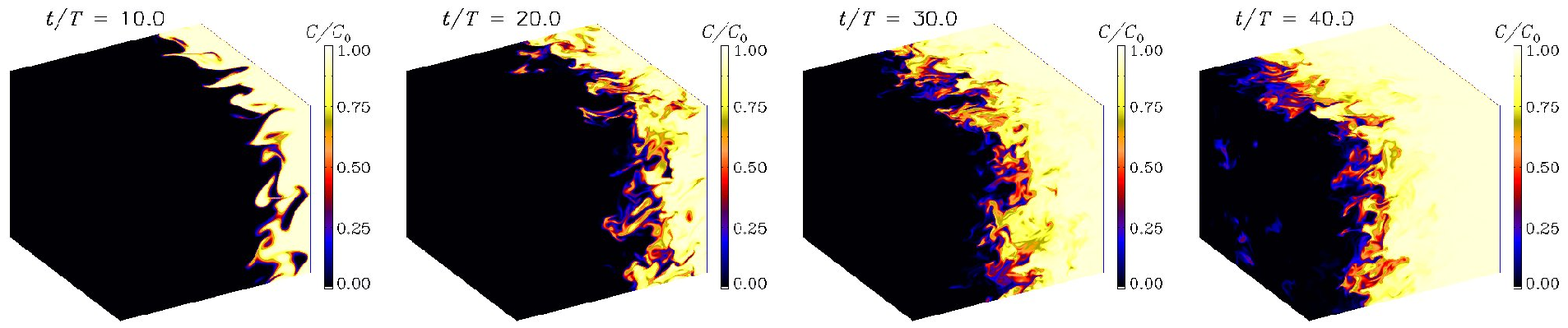}
\end{center}\caption[]{
Visualization of the concentration $C$ on the periphery of the box
at different times for Run~A1.
Here, $T=(\urms\kf)^{-1}$ is the turnover time.
}\label{256b}\end{figure*}

\section{DNS of the Fisher equation}

We now consider DNS of \Eq{dCdt_3D} where $\UU$ is obtained by solving
the Navier-Stokes equation for an isothermal gas with a forcing term
that is $\delta$ correlated in time.
The forcing function consists of plane waves whose
wave vector is random and its length is within a narrow window around
some mean forcing wavenumber $\kf$.
Since the gas is compressible and the density $\rho$ is not constant,
\Eq{dCdt_3D} now takes the form
\EQ
\frac{\partial C}{\partial t}+\UU\cdot\nab C
={C\over\tauc}\left(1-{C\over C_0}\right)
+\nab\cdot\left(\rho D\nab{C\over\rho}\right),
\label{dCdt_3Drho}
\EN
which we solve together with the momentum and continuity
equations,
\EQ
{\partial\UU\over\partial t}=-\UU\cdot\nab\UU-\cs^2\nab\ln\rho
+\ff+\FF_{\rm force},
\EN
\EQ
{\partial\rho\over\partial t}=-\nab\cdot\rho\UU,
\EN
where
$\FF_{\rm force}=\nu(\nabla^2\UU+\onethird\nab\nab\cdot\UU+2\SSSS\nab\ln\rho)$
is the viscous force, $\SSSS=\half\left[\nab\UU+(\nab\UU)^{\rm T}\right]
-\onethird\IIII\nab\cdot\UU$ is the traceless rate of strain tensor,
$\IIII$ is the unit matrix, $\nu$ is the molecular diffusivity,
and $\cs=\const$ is the isothermal sound speed.
The forcing function $\ff$ is of the form
\EQ
\ff(\xx,t)=\Reynolds\{N\ff_{\kk(t)}\exp[\ii\kk(t)\cdot\xx+\ii\phi(t)]\},
\EN
where $\xx$ is the position vector.
The wave vector $\kk(t)$ and the random phase
$-\pi<\phi(t)\le\pi$ change at every time step.
For the time-integrated forcing function to be independent
of the length of the time step $\delta t$, the normalization factor $N$
has to be proportional to $\delta t^{-1/2}$.
On dimensional grounds it is chosen to be
$N=f_0 c_{\rm s}(\kf c_{\rm s}/\delta t)^{1/2}$, where $f_0$ is a
nondimensional forcing amplitude.
The value of the coefficient $f_0$ is chosen such that the maxi\-mum Mach
number stays below about 0.5.
Here we choose $f_0=0.02$.
We force the system with nonhelical transversal waves,
\EQ
\ff_{\kk}=\left(\kk\times\eee\right)/\sqrt{\kk^2-(\kk\cdot\eee)^2},
\label{nohel_forcing}
\EN
where $\eee$ is an arbitrary unit vector not aligned with $\kk$;
note that $|\ff_{\kk}|^2=1$.

In the $x$ direction we use periodic boundary conditions for $\UU$
and $\rho$ and $\partial C/\partial x=0$ for $C$, while
we use periodic boundary conditions in the $y$ and $z$ directions.
The simulations were performed with the {\sc Pencil Code} \cite{PencilCode},
which uses sixth-order explicit finite differences in space and a
third-order accurate time stepping method \cite{BD02}.

For the calculations we use units where $k_1=\cs=\rho_0=1$.
However, most of the results are presented in an explicitly non-dimensional
form by normalizing with respect to relevant quantities such as the rms
velocity of the turbulence or the turnover time.
Our simulations are characterized by several non-dimensional parameters.
In addition to the values of Da and Pe, defined in \Eqs{256b}{Pe},
respectively, there is the Schmidt number, $\Sc=\nu/D$.
In those cases where the Damk\"ohler number was large, we had to increase
the value of $D$ in order to resolve the flame front.
This was done by decreasing $\Sc$ to values below unity.
The degree of scale separation is given by the ratio $\kf/k_1$.

\begingroup
\begin{table}[t!]
\centering
\caption{
Summary of the runs discussed in this paper.
}
\label{Tsum}
\begin{tabular}{ccrrrrccccc}
Run & $A$ & Re & Pe & Da &  Ka & $\kf/k_1$ &
$\sT/\uoneD$ & $\sT/\sL$ & $\uoneD/\sL$& $\ell/\ell_{\rm F}$ \\
\hline
A1&$1$&$117$&$117$&$  0.2$&$ 14.6$&$  5.0$&$ 0.97$&$ 7.50$&$ 7.75$&$ 15.0$\\
A2&$1$&$115$&$115$&$  0.5$&$  4.8$&$  5.1$&$ 1.60$&$ 7.03$&$ 4.38$&$ 26.3$\\
A3&$1$&$122$&$ 49$&$  1.6$&$  3.6$&$  5.1$&$ 2.33$&$ 3.77$&$ 1.61$&$ 30.4$\\
A4&$1$&$121$&$ 12$&$  4.8$&$  4.8$&$  5.1$&$ 3.55$&$ 1.63$&$ 0.46$&$ 26.3$\\
A5&$1$&$120$&$  4$&$ 15.9$&$  3.6$&$  5.1$&$ 7.29$&$ 1.16$&$ 0.16$&$ 30.3$\\
B1&$4$&$ 38$&$513$&$  1.6$&$  3.6$&$  1.6$&$ 1.88$&$ 9.76$&$ 5.19$&$ 98.9$\\
B2&$2$&$ 41$&$164$&$  4.9$&$  3.6$&$  1.6$&$ 2.21$&$ 3.69$&$ 1.67$&$ 98.8$\\
B3&$2$&$165$&$165$&$  4.9$&$  3.6$&$  1.6$&$ 2.52$&$ 4.22$&$ 1.67$&$ 98.9$\\
B4&$2$&$ 41$&$ 16$&$  4.9$&$ 36.4$&$  1.6$&$ 2.80$&$ 1.47$&$ 0.53$&$ 31.3$\\
B5&$2$&$ 40$&$ 16$&$ 50.3$&$  3.6$&$  1.6$&$ 7.25$&$ 1.19$&$ 0.16$&$ 98.9$\\
\end{tabular}
\end{table}
\endgroup

\section{Results}

\begin{figure}[t!]\begin{center}\includegraphics[width=\columnwidth]{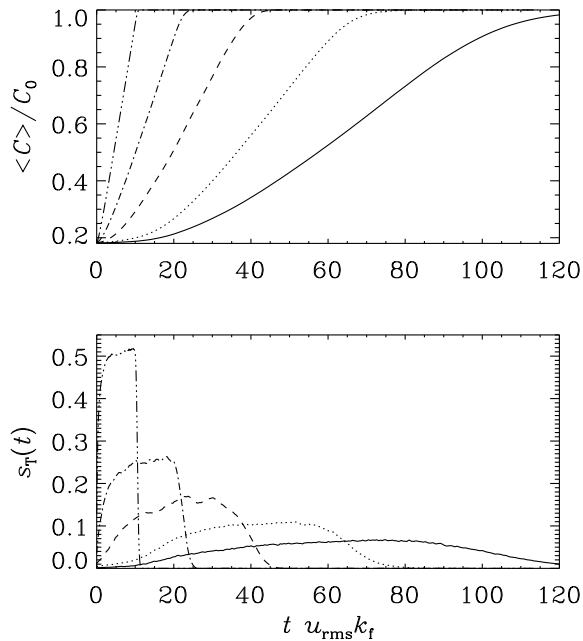}
\end{center}\caption[]{
Mean concentration and the instantaneous front speed as functions of time.
}\label{pcomp}\end{figure}

In the following we present results for uniform aspect ratio,
$A=L_x/L_y=1$ with $\kf/k_1=5$ (series~A), and $A=2$ or 4
with $\kf/k_1=1.6$ (series~B).
Our runs of series A and B are summarized in \Tab{Tsum}.
The resolution in the $y$ and $z$ directions is always $256^2$
meshpoints, but it is larger in the $x$ direction in 
runs where the aspect ratio $A$ is larger than unity.
In \Fig{256b} we show the concentration $C$ on the peri\-phe\-ry of the
computational domain at different times for $\tauc=3/\cs k_1$,
which corresponds to $\Da=0.5$; see \Tab{Tsum}.
One sees clearly how the front spreads and propagates in the negative
$x$ direction.
The front speed is determined in the same way as for the mean-field model,
i.e.\ using \Eq{vFmeasure}, except that $\meanC$ is computed from the actual $C$.
This can also be formulated as a volume integral,
\EQ
\vF(t)={1\over L_x L_y}{\dd\over\dd t}\int {C\over C_0}\,\dd V.
\EN
In \Fig{pcomp} we show examples of the evolution of the mean concentration
and the instantaneous front speed as functions of time for series~A.
The resulting ratios $\sT/\uoneD$, $\sT/\sL$, $\uoneD/\sL$, and
$\ell/\ell_{\rm F}$ are summarized in \Tab{Tsum} for series A and B.

\begin{figure}[t!]\begin{center}\includegraphics[width=\columnwidth]{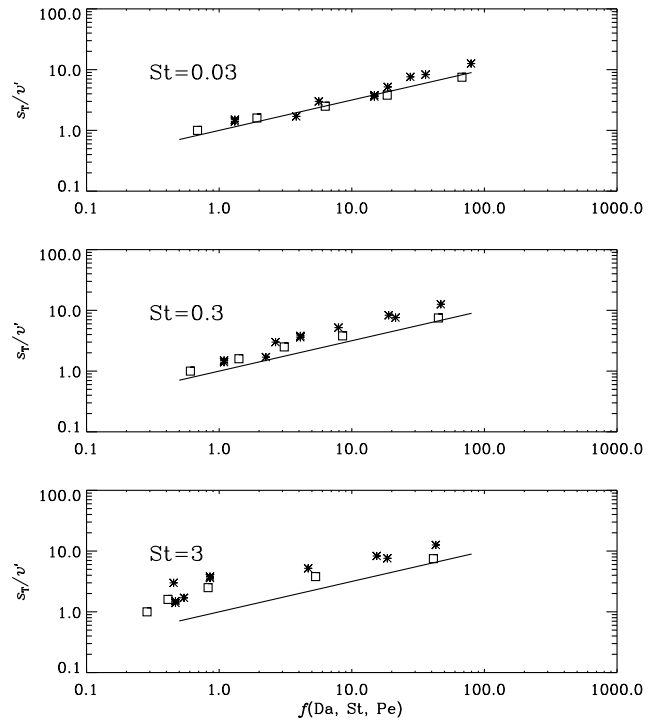}
\end{center}\caption[]{
Relative front speed as a function of $f$ for three values of St.
The squares indicate runs where the fluid is at rest and the front
is moving through the domain while the asterisk denote runs with
an inlet velocity chosen such that the front is approximately stationary
within the domain.
The solid line gives the theoretically expected result,
$\vF/\uoneD=f(\Da,\,\St,\,\Pe)^{1/2}$.
Note that the best agreement with the theoretical values is achieved for St=0.03.
}\label{pvF1}\end{figure}

In most of the cases considered in this paper,
the value of $\Pe$ is not in the asymptotic regime.
It might therefore be sensible to compare the relative front speed,
$\vF/\uoneD$ against the function $f(\Da,\,\St,\,\Pe)$.
This is done in \Fig{pvF1}, where we show the non-dimensional front speed,
$\vF/\uoneD$, versus $f(\Da,\,\St,\,\Pe)$, for three values of St
using values of Da and Pe, as evaluated from \Eqs{Da}{Pe}.
Surprisingly, the best fit is obtained for rather small values of St of 0.03.
This suggests that, for the present applications, the relevant value of $\tau$
is much smaller than in the case of a non-reacting passive scalar.

\begin{figure}[t!]\begin{center}\includegraphics[width=\columnwidth]{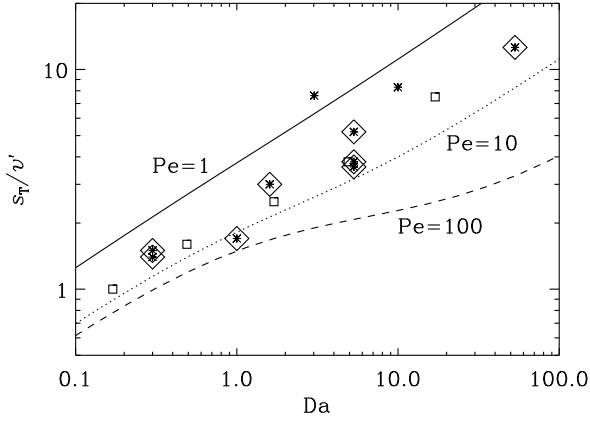}
\end{center}\caption[]{
Relative turbulent front speed versus Da.
The squares indicate runs where the fluid is at rest and the front
is moving through the domain while the asterisk denote runs with
an inlet velocity chosen such that the front is approximately stationary
within the domain.
For the latter, big asterisks denote cases where $\Pe>10$.
The lines give the theoretical expectations for $\St=0.03$
and Pe=1 (solid line), 10 (dotted), and 100 (dashed line).
}\label{psT}\end{figure}

\begin{figure}[t!]\begin{center}\includegraphics[width=\columnwidth]{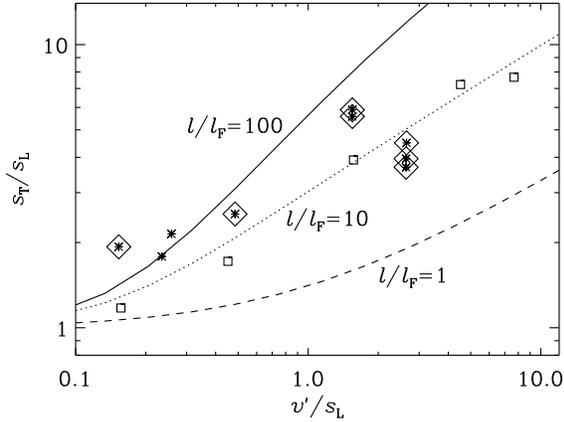}
\end{center}\caption[]{
Turbulent front speed versus turbulence intensity for
$\epsilon=0.1$ (solid line), 1 (dotted), and 10 (dashed)
using $\St=0.03$.
The squares indicate runs where the fluid is at rest and the front
is moving through the domain while the asterisk denote runs with
an inlet velocity chosen such that the front is approximately stationary
within the domain.
For the latter, big asterisks denote cases where $\Pe>10$.
The lines give the theoretical expectations.
}\label{psT_u1D}\end{figure}

Next, we plot $\sT/\uoneD$ versus Da for different values of Pe;
see \Fig{psT} using the previously inferred value St=0.03.
The data points from the DNS tend to lie between the curves for $\Pe=1$
and 10, even though most of the actual values of Pe are beyond Pe=10.
This too suggests some inconsistency between the DNS and the mean-field
description in terms of the telegraph equation.
Finally we plot the DNS results in a state diagram of $\sT/\sL$ versus
$\uoneD/\sL$ using St=0.03; see \Fig{psT_u1D}.
The data lie between the theoretical curves for $\ell/\ell_{\rm F}=10$ and
100, which is roughly in agreement with the values given in \Tab{Tsum}.

\section{Conclusions}

In the present work, the Fisher equation has served as a simple model
equation for front propagation in a turbulent flow.
The model has similarities with turbulent combustion, but is much simpler.
Nevertheless, it is clear that even this simple model harbors surprises
that one might have overlooked under more complex conditions.
Using three-dimensional simulations we have been able to compare with
the associated mean-field model.
For small Damk\"ohler numbers, the effective front speed can be approximated
by replacing the diffusivity by a turbulent value.
However, for Damk\"ohler numbers larger than unity, this simple procedure
fails, because it would suggest front speeds that exceed the characteristic
speed of the turbulent eddies.
A simple remedy is then to use a non-Fickian diffusion law for the
turbulent diffusion and to retain the time derivative in the expression
for the concentration flux.
Earlier work did already confirm the principal validity of this
approach and resulted in an estimate for the relevant relaxation time,
which is characterized by the Strouhal number.
The current work shows that the best fit to the simulation data can be
achieved with a Strouhal number that is as small as 0.03, which is about
100 times smaller than the earlier determined value
for passive scalar diffusion in forced turbulence.
This difference is connected with the presence of a reaction term
in the evolution equation for the passive scalar concentration.

\appendix
\section{Mean-field effect of the reaction term}
\label{MeanField}

In order to assess the effect of neglecting the reaction term in the
analysis presented above, we present now a simple mean-field theory
for the Fisher equation using the $\tau$ equation.
We start with the passive scalar equation with a reaction term
as given by \Eq{dCdt_3D}, split $C=\meanC+c$ and $\UU=\meanUU+\uu$
into mean and fluctuating parts, neglect the molecular diffusion term
for simplicity, and define the mean concentration flux
$\meanFFFF=\overline{\uu c}$ and the mean squared concentration,
$\meanHHH=\overline{c^2}$, so the equation for the mean concentration is
\EQ
{\partial\meanC\over\partial t}=
-\nab\cdot(\meanUU\,\meanC+\meanFFFF)
+{\meanC\over\tauc}\left(1-{\meanC\over C_0}\right)-{\meanHHH\over\tauc C_0},
\label{dmeanCdtH}
\EN
so the equation for the fluctuations is, to linear order in the fluctuations,
\EQ
{\partial c\over\partial t}=
-\nab\cdot(\meanUU c+\uu\meanC)
+{c\over\tauc}\left(1-{2\meanC\over C_0}\right)
+...
\EN
where the dots denote higher order terms for which we shall adopt a
general closure assumption.
Next, we derive evolution equations for $\meanFFFF$ and $\meanHHH$,
ignore a mean flow for simplicity, and assume $\nab\cdot\uu=0$, so we have
\EQA
{\partial\meanFFFF\over\partial t}&=&
-\tilde{\D}_{\rm t}\nab\meanC
+{\meanFFFF\over\tauc}\left(1-{2\meanC\over C_0}\right)
-{\meanFFFF\over\tau},
\label{dFdt}
\\
{\partial\meanHHH\over\partial t}&=&
-2\meanFFFF\cdot\nab\meanC
+2{\meanHHH\over\tauc}\left(1-{2\meanC\over C_0}\right)
-{\meanHHH\over\tau}.
\label{dHdt}
\ENA
In \Eqs{dFdt}{dHdt} we can write the last two terms as
$-\meanFFFF/\tau_{{\cal F}}$ and $-\meanHHH/\tau_{{\cal H}}$,
respectively, where
\EQA
{1\over\tau_{{\cal F}}(\meanC)}&=&{1\over\tau}-{1\over\tauc}
\left(1-{2\meanC\over C_0}\right),
\label{tauF}
\\
{1\over\tau_{{\cal H}}(\meanC)}&=&{1\over\tau}-{2\over\tauc}
\left(1-{2\meanC\over C_0}\right).
\ENA
On sufficiently long time scales we may ignore the time derivatives
in \Eqs{dFdt}{dHdt}, so we arrive at closed expressions for
$\meanFFFF$ and $\meanHHH$, that we insert into \Eq{dmeanCdtH}, so we obtain
\EQ
{\partial\meanC\over\partial t}+\UU_{\rm c}\cdot\nab\meanC=
{\meanC\over\tauc}\left(1-{\meanC\over C_0}\right)
+\DT\nabla^2\meanC,
\EN
where $\UUc(\meanC)$ is a new effective advection speed and
$\DT=\D+\Dt$ is again the sum of turbulent and microscopic diffusivities with
\EQ
\UUc(\meanC)=2\Dt{\tau_{{\cal H}}\over\tauc}{\nab\meanC\over C_0}
,\quad
\Dt(\meanC)=\tau_{{\cal F}}(\meanC)\,{v'}^2.
\label{DtCmean}
\EN
One may expect that the term $\UUc$ slows
down the propagation speed of the front, because it is directed up the
concentration gradient.
Note that the sign of the $\UUc$ term is opposite
to that of a similar term in the
so-called $G$ equation \cite{Pet99,Wil85} of turbulent front propagations,
which is however not an equation for the flame brush, but for the detailed
position of the wrinkled flame front (at $G=0$) with an advection speed that
is given by $\uu-\sL\nnn$, where $\nnn=\nab G/|\nab G|$ is a unit vector
normal to the flame front, but it enters with a minus sign and thus
corresponds to an enhanced speed down the gradient of $G$.
However, by solving \Eq{dmeanCdtH} with \Eqs{dFdt}{dHdt},
it turns that when the $\meanHHH$ term is included, it accelerates the front;
see \Tab{Tsum2}.
Note also that the coefficient $\Dt$ is reduced and can even become
negative in the unstable part of the front where $C=0$ (or at least
$C<C_0/2$); see \Eqs{tauF}{DtCmean}.
In that case our expression for turbulent diffusion becomes invalid
and one has to include higher order derivatives that would
guarantee stability at small length scales.

\begingroup
\begin{table}[t!]
\centering
\caption{
Dependence of $\sT/\uoneD$ without and with $\meanHHH$
in a model for $\Pe=10$.
Note the slight increase of $\sT/\uoneD$ when $\meanHHH$
compared to the case where it is neglected.
}
\label{Tsum2}
\begin{tabular}{ccc}
~~Da~~ & ~~$\sT/\uoneD$(without $\meanHHH$)~~ & ~~$\sT/\uoneD$(with $\meanHHH$)~~ \\
\hline
  0.10 &  0.25 &  0.25 \\
  0.30 &  0.44 &  0.47 \\
  0.50 &  0.59 &  0.65 \\
  0.61 &  0.66 &  0.73 \\
\end{tabular}
\end{table}
\endgroup

\acknowledgments
We acknowledge the allocation of computing resources provided by the
Swedish National Allocations Committee at the Center for
Parallel Computers at the Royal Institute of Technology in
Stockholm and the National Supercomputer Centers in Link\"oping.
This work was supported in part by
the European Research Council under the AstroDyn Research Project 227952,
the Swedish Research Council grant 621-2007-4064 and the European 
Community's Seventh Framework Programme (FP7/2007-2013) under grant 
agreement nr 211971 (The DECARBit project) (NELH).


\end{document}